\documentstyle[pre,aps]{revtex}

\draft

\begin{document}
\title{Scaling for domain growth in the Ising model with competing dynamics}
\author{Zhi-Feng Huang,$^1$ Bing-Lin Gu,$^{1,2}$ and Yun Tang$^3$}
\address{$^1$Center for Advanced Study, Tsinghua University, Beijing 100084, People's%
\\
Republic of China\\
$^2$Department of Physics, Tsinghua University, Beijing 100084, People's\\
Republic of China\\
$^3$Department of Applied Mathematics, Tsinghua University, Beijing 100084,\\
People's Republic of China}
\date{\today}
\maketitle

\begin{abstract}
We study the domain growth of the one-dimensional kinetic Ising model under
the competing influence of Glauber dynamics at temperature $T$ and Kawasaki
dynamics with a configuration-independent rate. The scaling of the structure
factor is shown to have the form for nonconserved dynamics with the
corrections arising from the spin-exchange process, i.e., $%
S(k,t)=Lg_0(kL,t/\tau )+g_1(kL,t/\tau )+\cdots $, and the corresponding
scaling functions are calculated analytically. A correction to the Porod law
at zero temperature is also given.
\end{abstract}


The kinetic Ising models have been widely studied to understand the
far-from-equilibrium phenomena such as the nonconserved Glauber model \cite
{Glauber63} and the conserved Kawasaki model \cite{Kawasaki72}. In recent
years, the systems with more complex mechanisms, e.g., the Ising models with
competing Glauber and Kawasaki dynamics \cite{DFL85,TO89}, have been
introduced to investigate the basic questions of nonequilibrium phase
transitions and critical phenomena. For these models with competing
dynamics, much of the interest has been focused on the properties of
nonequilibrium steady state, including the nonequilibrium phase diagrams 
\cite{TO89,GGML87,GF97} and the critical exponents \cite{GF97,WL88}.
However, the problem of domain growth, which describes the behavior of the
system quenched into an ordered phase from a high-temperature initial state
and is important in understanding the dynamics of nonequilibrium processes,
has received relatively little attention.

Our interest here is in the study of domain growth for the Ising system with
competing dynamics. It is well known \cite{GMS83,Bray94} that in the late
stage regime, the nonequilibrium process of domain growth exhibits dynamic
scaling behaviors and the scaling forms of the equal-time pair correlation
function $C(r,t)$ and its Fourier transformation, the structure factor $%
S(k,t)$, are given by $C(r,t)=f\left( r/L(t)\right) $ and $%
S(k,t)=L(t)^dg\left( kL(t)\right) $, respectively, where $L(t)$ is the
single length scale characterizing the domain structure and $d$ is the
spatial dimensionality. The corrections to the above scaling arising from
the nonscaling initial condition were determined recently at zero
temperature \cite{BRC98}. However, the direct demonstration of scaling and
the exact calculation of the scaling functions are still lacking except in
some simple models \cite{BRC98,Huang97,Bray90,AF90}. Thus it is of interest
to give a direct and exact calculation of scaling behaviors for the system
with more complex dynamics.

The system considered here is the one-dimensional Ising model evolving by a
combination of the spin-flip Glauber process at temperature $T$ and the
spin-exchange Kawasaki process with a configuration-independent rate that
occurs as if the system were at infinite temperature \cite{DFL85,DRS89}. In
this work we study the domain growth of this soluble model following a
sudden quench from an initial high temperature to a final temperature $T$
and calculate the corresponding scaling properties analytically.

The Hamiltonian for this one-dimensional system is given by 
\begin{equation}
{\cal H}=-J\sum\limits_{i=1}^N\sigma _i\sigma _{i+1},  \label{HamiltonP}
\end{equation}
where $\sigma _i=\pm 1$ is the Ising spin variable. The configuration of the
system evolves with time via the combination of Glauber dynamics with the
spin-flip rate $W_i^{(1)}=(2\tau _1)^{-1}[1-\gamma \sigma _i\left( \sigma
_{i-1}+\sigma _{i+1}\right) /2],$ where $\gamma =\tanh 2K$ with $K=J/k_BT$,
and Kawasaki dynamics with the configuration-independent spin-exchange rate $%
W_{ii+1}^{(2)}=(2\tau _2)^{-1}\left( 1-\sigma _i\sigma _{i+1}\right) .$ The
equations for the expectation value of the spin and the equal-time pair
correlation function have been derived \cite{DRS89} and the latter is
written as 
\begin{eqnarray}
\tau ^{\prime }dC(r,t)/dt &=&-2C(r,t)+\gamma ^{\prime }\left[
C(r-1,t)+C(r+1,t)\right] \quad (r\geq 2),  \nonumber \\
\tau ^{\prime }dC(1,t)/dt &=&-\frac{2+\kappa }{1+\kappa }C(1,t)+\gamma
^{\prime }C(2,t)+\gamma ^{\prime }-\frac \kappa {1+\kappa },  \nonumber \\
C(0,t) &=&1  \label{CorrelP}
\end{eqnarray}
provided the initial probability distribution is translationally invariant,
where the pair correlation function $C(r,t)=\langle \sigma _i(t)\sigma
_{i+r}(t)\rangle $, $\kappa =2\tau _1/\tau _2$, $\tau ^{\prime }=\tau
_1/(1+\kappa )$, and $\gamma ^{\prime }=(\gamma +\kappa )/(1+\kappa )$.

From Eq. (\ref{CorrelP}) the scaling results for domain growth can be
calculated directly following the method given by Bray \cite{Bray90}. By
Fourier transforming in space, the equation for the structure factor $S(k,t)$
is obtained, 
\begin{eqnarray}
\tau ^{\prime }dS(k,t)/dt &=&-\gamma _kS(k,t)+A(t)-\frac{2\kappa }{1+\kappa }%
n(t)\cos k,  \nonumber \\
\frac 1N\sum\limits_kS(k,t) &=&1,  \label{StructP}
\end{eqnarray}
where $\gamma _k=2(1-\gamma ^{\prime }\cos k)$, $A(t)=\sum_k\gamma
_kS(k,t)/N $, and $n(t)=[1-C(1,t)]/2$ denotes the average wall density. This
equation can be solved by a Laplace transformation in time via $\bar{S}%
(k,s)=\int_0^\infty dt\exp (-st)S(k,t).$ In the scaling regime of late times
after the quench, the characteristic domain size $L(t)$ is known to obey the
growth law $L(t)\sim t^x$, where the growth exponent $x$ is $1/2$ and $1/3$
for systems with nonconserved and conserved scalar order parameters,
respectively \cite{GMS83,Bray94}. Its inverse, the average wall density,
then satisfies $n(t)=bt^{-x}/2=bt^{\nu -1}/2$ with $0<\nu <1$, as well as a
certain coefficient $b$, and the corresponding Laplace transform is $%
n(s)=b\Gamma (\nu )s^{-\nu }/2$. Here we assume that the above scaling forms
of the domain size and the wall density remain for this model with competing
dynamics, which will be justified {\it a posteriori}. Thus, on the condition
that the initial state contains no long-range order, in the scaling limit ($%
s\rightarrow 0$, $k\rightarrow 0$ with $s/k^{1/x}$ arbitrary) we have 
\begin{eqnarray}
\overline{S}(k,s) &=&\left\{ (\tau ^{\prime }s+\kappa ^{\prime
2})^{1/2}\left[ 2s^{-1}-\frac \kappa {1+\kappa }\left( 1+\frac{\kappa
^{\prime 2}}2\right) b\Gamma (\nu )s^{-\nu }\right] \right.  \nonumber \\
&&\left. +\frac \kappa {1+\kappa }\frac{k^2+\kappa ^{\prime 2}}2b\Gamma (\nu
)s^{-\nu }\right\} \left( \tau ^{\prime }s+k^2+\kappa ^{\prime 2}\right)
^{-1}  \label{LaplaceS}
\end{eqnarray}
when domains grow at small but nonzero temperature $T$. Here $\kappa
^{\prime }=\xi ^{-1}\simeq 2\exp (-2K)/\sqrt{1+\kappa }$, where $\xi $ is
the correlation length for this system with competing dynamics \cite{DRS89}.

Using the inverse Laplace transform on Eq. (\ref{LaplaceS}) yields 
\begin{eqnarray}
S(k,t) &=&2\left( \frac t{\pi \tau ^{\prime }}\right) ^{1/2}\frac 1{%
k^2+\kappa ^{\prime 2}}\int_0^1dy~y^{-1/2}\left[ \kappa ^{\prime 2}\exp
\left( -\frac{\kappa ^{\prime 2}t}{\tau ^{\prime }}y\right) +k^2\exp \left( -%
\frac{k^2+\kappa ^{\prime 2}}{\tau ^{\prime }}t+\frac{k^2t}{\tau ^{\prime }}%
y\right) \right]  \nonumber \\
&&+\frac{\kappa b}{1+\kappa }\left\{ \left( \frac 1{\pi \tau ^{\prime }}%
\right) ^{1/2}\left( 1+\frac{\kappa ^{\prime 2}}2\right) t^{\nu -1/2}\left[
-B\left( \frac 12,\nu \right) \ _1F_1\left( \nu ;\frac 12+\nu ;\frac{\kappa
^{\prime 2}t}{\tau ^{\prime }}\right) \exp \left( -\frac{\kappa ^{\prime 2}t%
}{\tau ^{\prime }}\right) \right. \right.  \nonumber \\
&&\left. +\frac 1\nu \frac{k^2t}{\tau ^{\prime }}\int_0^1dy~y^{-1/2}(1-y)^%
\nu \ _1F_1\left( \nu ;1+\nu ;\frac{k^2+\kappa ^{\prime 2}}{\tau ^{\prime }}%
t(1-y)\right) \exp \left( -\frac{k^2+\kappa ^{\prime 2}}{\tau ^{\prime }}t+%
\frac{k^2t}{\tau ^{\prime }}y\right) \right]  \nonumber \\
&&\left. +\frac{k^2+\kappa ^{\prime 2}}{2\tau ^{\prime }\nu }t^\nu \
_1F_1\left( \nu ;1+\nu ;\frac{k^2+\kappa ^{\prime 2}}{\tau ^{\prime }}%
t\right) \exp \left( -\frac{k^2+\kappa ^{\prime 2}}{\tau ^{\prime }}t\right)
\right\} ,  \label{ScalingTP}
\end{eqnarray}
where $B(x,y)$ is the Beta function and $_1F_1(\alpha ;\beta ;z)$ is the
degenerate hypergeometric function \cite{GR80}. It is shown from Eq. (\ref
{ScalingTP}) that a scaling solution requires $L(t)\sim t^{1/2}$, i.e., $\nu
=1/2$, and then the equal-time structure factor has the scaling form 
\begin{equation}
S(k,t)=t^{1/2}g_0(k^2t,\kappa ^{\prime 2}t)+g_1(k^2t,\kappa ^{\prime
2}t)+t^{-1/2}g_2(k^2t,\kappa ^{\prime 2}t)+t^{-1}g_3(k^2t,\kappa ^{\prime
2}t),  \label{SformTP}
\end{equation}
with the scaling functions 
\begin{eqnarray}
g_0(k^2t,\kappa ^{\prime 2}t) &=&2\left( \frac 1{\pi \tau ^{\prime }}\right)
^{1/2}\frac 1{k^2+\kappa ^{\prime 2}}\int_0^1dy~y^{-1/2}\left[ \kappa
^{\prime 2}\exp \left( -\frac{\kappa ^{\prime 2}t}{\tau ^{\prime }}y\right)
+k^2\exp \left( -\frac{k^2+\kappa ^{\prime 2}}{\tau ^{\prime }}t+\frac{k^2t}{%
\tau ^{\prime }}y\right) \right] ,  \label{function0} \\
g_1(k^2t,\kappa ^{\prime 2}t) &=&\frac{\kappa b}{1+\kappa }\left( \frac 1{%
\pi \tau ^{\prime }}\right) ^{1/2}\left[ -\pi \ _1F_1\left( \frac 12;1;\frac{%
\kappa ^{\prime 2}t}{\tau ^{\prime }}\right) \exp \left( -\frac{\kappa
^{\prime 2}t}{\tau ^{\prime }}\right) +2\frac{k^2t}{\tau ^{\prime }}%
\int_0^1dy~y^{-1/2}(1-y)^{1/2}\right.  \nonumber \\
&&\left. \times \ _1F_1\left( \frac 12;\frac 32;\frac{k^2+\kappa ^{\prime 2}%
}{\tau ^{\prime }}t(1-y)\right) \exp \left( -\frac{k^2+\kappa ^{\prime 2}}{%
\tau ^{\prime }}t+\frac{k^2t}{\tau ^{\prime }}y\right) \right] ,
\label{function1} \\
g_2(k^2t,\kappa ^{\prime 2}t) &=&\frac{\kappa b}{1+\kappa }\frac{k^2+\kappa
^{\prime 2}}{\tau ^{\prime }}t\ _1F_1\left( \frac 12;\frac 32;\frac{%
k^2+\kappa ^{\prime 2}}{\tau ^{\prime }}t\right) \exp \left( -\frac{%
k^2+\kappa ^{\prime 2}}{\tau ^{\prime }}t\right) ,  \label{function2} \\
g_3(k^2t,\kappa ^{\prime 2}t) &=&\frac{\kappa b}{1+\kappa }\left( \frac 1{%
\pi \tau ^{\prime }}\right) ^{1/2}\frac{\kappa ^{\prime 2}t}2\left[ -\pi \
_1F_1\left( \frac 12;1;\frac{\kappa ^{\prime 2}t}{\tau ^{\prime }}\right)
\exp \left( -\frac{\kappa ^{\prime 2}t}{\tau ^{\prime }}\right) +2\frac{k^2t%
}{\tau ^{\prime }}\int_0^1dy~y^{-1/2}(1-y)^{1/2}\right.  \nonumber \\
&&\left. \times \ _1F_1\left( \frac 12;\frac 32;\frac{k^2+\kappa ^{\prime 2}%
}{\tau ^{\prime }}t(1-y)\right) \exp \left( -\frac{k^2+\kappa ^{\prime 2}}{%
\tau ^{\prime }}t+\frac{k^2t}{\tau ^{\prime }}y\right) \right] .
\label{function3}
\end{eqnarray}
Thus, from this scaling result of the structure factor it is shown that the
domain size $L(t)$ scales as $t^x$ and, correspondingly, the wall density $%
n(t)$ has the scaling form $t^{-x}$, which is consistent with the assumption
made above before the calculation. Here the growth exponent $x$ is $1/2$ for
the competing dynamics, the same as that for the (single) nonconserved
Glauber system.

For $\kappa ^{\prime 2}t\gg 1$, one can obtain the asymptotic result of $%
S(k,t)$, that is, 
\begin{equation}
S(k,t)\longrightarrow \frac{2\kappa ^{\prime }}{k^2+\kappa ^{\prime 2}}+%
\frac{\kappa b}{1+\kappa }\frac{1-\kappa ^{\prime }}2t^{-1/2},
\label{asymTP}
\end{equation}
while for $\kappa ^{\prime 2}t\rightarrow 0$, the scaling result for quench
to $T=0$, which is the main interest in the one-dimensional system, is given
by 
\begin{eqnarray}
S(k,t) &=&2\left( \frac t{\pi \tau ^{\prime }}\right)
^{1/2}\int_0^1dy~y^{-1/2}\exp \left[ -\frac{k^2t}{\tau ^{\prime }}%
(1-y)\right]  \nonumber \\
&&+\frac{\kappa b}{1+\kappa }\left\{ \left( \frac 1{\pi \tau ^{\prime }}%
\right) ^{1/2}\left[ -\pi +2\frac{k^2t}{\tau ^{\prime }}%
\int_0^1dy~y^{-1/2}(1-y)^{1/2}\ _1F_1\left( \frac 12;\frac 32;\frac{k^2t}{%
\tau ^{\prime }}(1-y)\right) \right. \right.  \nonumber \\
&&\left. \left. \times \exp \left( -\frac{k^2t}{\tau ^{\prime }}(1-y)\right)
\right] +t^{-1/2}\frac{k^2t}{\tau ^{\prime }}\ _1F_1\left( \frac 12;\frac 32;%
\frac{k^2t}{\tau ^{\prime }}\right) \exp \left( -\frac{k^2t}{\tau ^{\prime }}%
\right) \right\} ,  \label{Scaling0P}
\end{eqnarray}
which is of the scaling form 
\begin{equation}
S(k,t)=t^{1/2}g_0(k^2t)+g_1(k^2t)+t^{-1/2}g_2(k^2t).  \label{Sform0P}
\end{equation}
For $k^2t\gg 1$, Eq. (\ref{Scaling0P}) becomes 
\begin{equation}
S(k,t)\longrightarrow 2(\pi \tau ^{\prime }t)^{-1/2}k^{-2}+\frac{\kappa b}{%
2(1+\kappa )}t^{-1/2},  \label{asym0P}
\end{equation}
which yields a correction to the Porod law \cite{Porod82}.

It should be noted that the scaling forms of the structure factor (\ref
{SformTP}) at nonzero temperature and Eq. (\ref{Sform0P}) at zero
temperature can be rewritten as

\begin{equation}
S(k,t)=Lg_0(kL,t/\tau )+g_1(kL,t/\tau )+L^{-1}g_2(kL,t/\tau
)+L^{-2}g_3(kL,t/\tau ),  \label{FormST}
\end{equation}
with $\tau =1/\kappa ^{\prime 2}=\xi ^2$ and $L\sim t^{1/2}$, and 
\begin{equation}
S(k,t)=Lg_0(kL)+g_1(kL)+L^{-1}g_2(kL),  \label{FormS0}
\end{equation}
respectively. By Fourier transforming, the corresponding scaling of the
equal-time pair correlation function is obtained 
\begin{equation}
C(r,t)=f_0(r/L,t/\tau )+L^{-1}f_1(r/L,t/\tau )+L^{-2}f_2(r/L,t/\tau
)+L^{-3}f_3(r/L,t/\tau ),  \label{FormFT}
\end{equation}
at nonzero temperature and 
\begin{equation}
C(r,t)=f_0(r/L)+L^{-1}f_1(r/L)+L^{-2}f_2(r/L)  \label{FormF0}
\end{equation}
at $T=0$. In the expressions of these scaling forms, the first leading term
exhibits the scalings for the nonconserved dynamics in one dimension \cite
{Bray94,Bray90} associated with the spin-flip Glauber process, while the
other terms are the corrections to scaling that arise from the spin-exchange
Kawasaki process, which can be shown from the expressions of the scaling
functions. In formulas (\ref{function1})--(\ref{function3}) for quench to
nonzero temperature, the correction functions $g_1$, $g_2$, and $g_3$ vary
directly as $\kappa /(1+\kappa )$ and vanish if the contribution of the
spin-exchange process is negligible or, equivalently, $\kappa \rightarrow 0$%
. Consequently, the scaling form $S(k,t)=t^{1/2}g_0(k^2t,\kappa ^{\prime
2}t) $ with the scaling function $g_0(x,y)$ and the asymptotic result $%
S(k,t)\rightarrow 2\kappa ^{\prime }/(k^2+\kappa ^{\prime 2})$ for $\kappa
^{\prime 2}t\gg 1$ obtained by Bray \cite{Bray90} for the one-dimensional
Ising model with the (single) Glauber dynamics are recovered from Eqs. (\ref
{SformTP})--(\ref{asymTP}). Similar results can be obtained for zero
temperature, that is, one can recover the well known scaling form $%
S(k,t)=t^{1/2}g_0(k^2t)$ and the Porod law $S(k,t)\sim k^{-2}$ for Glauber
Ising chain \cite{Bray90,AF90} from Eqs. (\ref{Scaling0P})--(\ref{asym0P}).
Moreover, from Eqs. (\ref{FormFT}) and (\ref{FormF0}) we note that the
leading correction to the nonconserved scaling form of the pair correlation
function is $L^{-1}f_1(r/L)$ here, originating from the spin-exchange
process, while in the absence of this process, associated with departure of
the initial condition from the scaling morphology, it is $L^{-4}f_1(r/L)$,
according to the recent work of Bray {\it et al.} \cite{BRC98}.

From the above scaling solution no correction to scaling of the domain size $%
L(t)$ and the wall density $n(t)$ is shown, as expected before the
calculation. However, if assuming the correction to scaling of $L(t)$ and $%
n(t)$ before solving the evolution equation (\ref{StructP}), e.g., $n(t)\sim
b_1t^{-x_1}+b_2t^{-x_2}+\cdots $ with $0<x_1<x_2<\cdots $, we find that the
scaling solution similar to Eqs. (\ref{SformTP}) and (\ref{FormST}) cannot
be obtained.

In summary, we have shown that for this Ising system with competing Glauber
and Kawasaki dynamics, the spin-exchange process with a
configuration-independent rate has the effect of corrections to scaling,
while the leading contribution to scaling is of the form for the
nonconserved Glauber dynamics, as shown in Eqs. (\ref{ScalingTP}) and (\ref
{Scaling0P}). More work is needed for a further understanding of these
results.

Z.F.H. acknowledges helpful discussions with Guang-Ming Zhang.

\end{document}